# Hardware Acceleration of Kolmogorov–Arnold Network (KAN) for Lightweight Edge Inference

Wei-Hsing Huang[*,1], Jianwei Jia[*,1], Yuyao Kong[1], Faaiq Waqar[1], Tai-Hao Wen[2], Meng-Fan Chang[2,3], Shimeng Yu[1]
[1]Georgia Institute of Technology, Atlanta, GA, [2]National Tsing Hua University, Hsinchu, Taiwan,
[3]TSMC Corporate Research, Hsinchu, Taiwan, [*]Equal Contribution
Email: shimeng.yu@ece.gatech.edu

*Abstract*—Recently, a novel model named Kolmogorov-Arnold Networks (KAN) has been proposed with the potential to achieve the functionality of traditional deep neural networks (DNNs) using orders of magnitude fewer parameters by parameterized B-spline functions with trainable coefficients. However, the B-spline functions in KAN present new challenges for hardware acceleration. Evaluating the B-spline functions can be performed by using look-up tables (LUTs) to directly map the B-spline functions, thereby reducing computational resource requirements. However, this method still requires substantial circuit resources (LUTs, MUXs, decoders, etc.). *For the first time*, this paper employs an algorithm-hardware co-design methodology to accelerate KAN. The proposed algorithm-level techniques include Alignment-Symmetry and PowerGap KAN hardware aware quantization, KAN sparsity aware mapping strategy, and circuit-level techniques include N:1 Time Modulation Dynamic Voltage input generator with analog-CIM (ACIM) circuits. The impact of non-ideal effects, such as partial sum errors caused by the process variations, has been evaluated with the statistics measured from the TSMC 22nm RRAM-ACIM prototype chips. With the best searched hyperparameters of KAN and the optimized circuits implemented in 22 nm node, we can reduce hardware area by 41.78x, energy by 77.97x with 3.03% accuracy boost compared to the traditional DNN hardware.

## 1 INTRODUCTION

Contemporary DNN models with ever-increasing parameter counts hinder edge deployment, limiting privacy-sensitive, real-time detection and resource-constrained edge applications. DNNs, such as multi-layer perceptron (MLP), convolutional neural network (CNN), large language model (LLM), and other architectures, typically employ fixed activation functions and learnable weights [1][2]. The recently proposed Kolmogorov-Arnold Networks (KAN) [3], inspired by the Kolmogorov-Arnold theorem [4][5], replaces traditional multi-layer perceptron (MLP)'s linear weights with B-spline functions B(X) and employs trainable activation functions. This architecture offers improved interpretability and has been shown to achieve or surpass traditional DNNs performance with smaller network size as shown in Fig. 1 [3][6]. Consequently, KAN architectures show potential to replace traditional DNN backbones that use fixed activation functions and learnable weights in complex

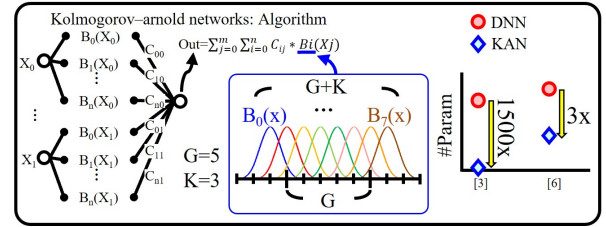

Fig. 1. Introduction of KAN and its potential for parameter reduction.

DNN models, potentially reducing the size of large models, including LLM, and facilitating their edge deployment.

However, KAN operation requires B-spline function computation. While mathematical definitions involving recursive methods [7] can evaluate B-spline functions, computational requirements increase significantly with higher-order k. An edge-friendly alternative employs look-up tables (LUTs) for direct B-spline function mapping, simplifying hardware implementation and reducing computational demands. Despite these advantages, this approach still necessitates significant circuit resources (LUTs, MUXs, decoders, etc.) as shown in Fig. 2.

Furthermore, KAN, like traditional MLPs, involves extensive parallel MAC operations. The von Neumann bottleneck in conventional architectures leads to inefficiencies. Compute-in-Memory (CIM) [8], a non-von Neumann architecture, addresses this issue. CIM variants include digital-CIM (DCIM) [9], SRAM analog-CIM (ACIM) [10][11], and RRAM-ACIM [12][13], etc. While DCIM and SRAM-ACIM offer higher accuracy than RRAM-ACIM, large SRAM cell sizes limit on-chip capacity, and high standby power consumption is undesirable for edge devices. The paper examines RRAM-ACIM acceleration for KAN. However, note that the proposed algorithm-level optimizations are hardware-agnostic.

Despite RRAM-ACIM's advantages for edge deployment, including low standby power, it encounters the challenges as shown in Fig. 2. The process-temperature-voltage (PVT) variations are key concerns for ACIM. The increasing array sizes and technology scaling exacerbate IR-drop [14] issues due to increased bit line resistance, hampering inference accuracy. Moreover, current mainstream CIM input methods include binary input with multi-cycle and multi-voltage level input within one cycle or time delay with multi-bit input. Binary input offers higher accuracy but increases latency and limits TOPS/W. While multi-voltage level input approach achieves favorable TOPS/W ratios and low latency, the constrained VDD range renders inputs susceptible to noise, thereby compromising accuracy. Time-delay multi-bit inputs offer better noise resilience but at the cost of increased latency.

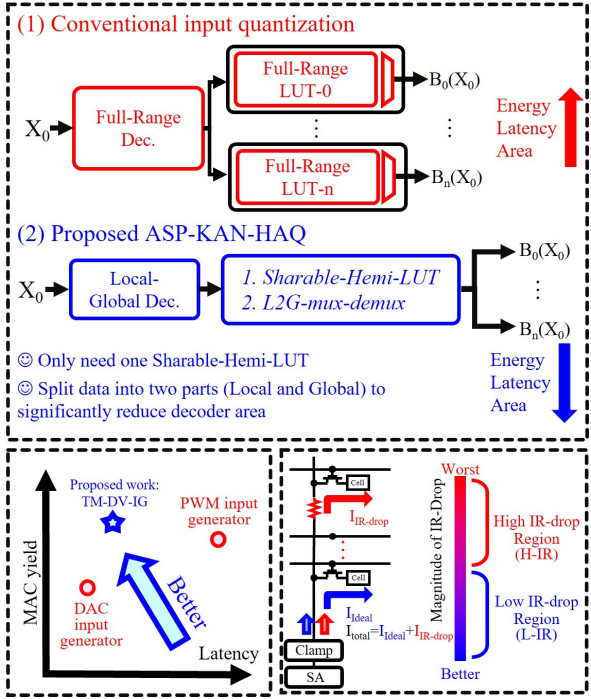

Fig. 2. Challenges hindering low-power high accuracy edge applications.

In this work, we aim to co-optimize algorithm and circuit, reducing area, power, and latency while increasing the accuracy of ACIM computation for KAN. Our key contributions include:

- We propose an Alignment-Symmetry and PowerGap KAN hardware aware quantization that, **for the first time**, investigates the interaction between quantization grid and knot grid in KAN. The proposed method significantly minimizes the cost of LUTs, MUXs, decoders for B(X) function.
- We propose N:1 Time Modulation Dynamic Voltage input generator, a mixed time-voltage input generator to reduce area and latency as well as increase MAC yield for multi-level input generator compared to both traditional pure voltage-based DAC and pure latency-based multi-level input generator.
- We propose a KAN sparsity aware weight mapping strategy to reduce the IR-drop issue while operating ACIM to improve the inference accuracy.
- We present KAN hyperparameter optimization framework with a modified version of NeuroSim [17], KAN-NeuroSim. KAN-NeuroSim would search the best hyperparameter of G in KAN for Alignment-Symmetry and PowerGap KAN hardware aware quantization and N:1 Time Modulation Dynamic Voltage input generator for RRAM-ACIM.

## 2  BACKGROUND

### 2.1 KAN: Kolmogorov-Arnold Networks

Unlike MLP, which is inspired by the universal approximation theorem, KAN derives from the Kolmogorov-Arnold representation theorem. Specifically, each KAN layer is defined by equation (1)-(2), where b(x), originally composed of SiLU, introduces residual structure.

$$\phi(x) = w_b b(x) + w_s \text{spline}(x) \quad (1)$$
$$\text{spline}(x) = \sum_i c_i B_i(x) \quad (2)$$
$$\phi(x) = w_b b(x) + \sum_i c_i' B_i(x) \quad (3)$$

As illustrated in Fig. 1, G denotes the grid size, while K represents the order of B-splines. The total number of $B_i(x)$ is K+G, where in this example, K=3 and G=5. In our implementation, we replace SiLU with ReLU for improved hardware efficiency without accuracy loss. Since $w_b b(x)$ can be accelerated using traditional RRAM-ACIM, our focus is on accelerating $w_s \text{spline}(x)$ computation. In our implementation, $w_s$ is multiplied with $c_i$ and becomes $c_i'$, then is quantized to 8-bit, transforming the formula to equation (3). Once $B_i(x)$ is obtained, $c_i'$ can be stored in RRAM-ACIM, with $B_i(x)$ input via WL for parallel MAC operations. Direct LUT mapping of $B_i(x)$ is more suitable for edge devices compared to recursive evaluation of B-spline functions. In addition, due to KAN's utilization of uniform nodes, each B(X) represents the same function across different knot grid intervals. This uniformity creates the potential for implementing a shared LUT across all B(X) functions. However, as shown in Fig. 2, conventional quantization methods [15] misalign knot and quantization grids, resulting in distinct x-y correspondences for each $B_i(x)$. This necessitates individual LUTs, MUXs, and decoders for each $B_i(x)$ during edge implementation, incurring significant energy and area overhead. A straightforward solution is to employ non-programmable LUTs, thereby reducing LUT area. However, this approach compromises system flexibility, such as the ability to dynamically adjust B(X) precision based on application requirements. Moreover, as K and G increase, the number of $B_i(x)$ functions grows, thereby impeding the scalability of conventional quantization methods for more complex KAN networks (higher K and G) on edge devices. Section 3.1 presents the proposed Alignment-Symmetry and PowerGap KAN hardware aware quantization to address this issue.

### 2.2 Compute-in-memory

CIM architectures utilize various embedded memory technologies, including SRAM, eDRAM, and emerging non-volatile memories such as RRAM. Each type of memory offers distinct advantages and faces different challenges. For instance, SRAM-based CIM benefits from leading-edge node technology and fast access times but suffers from high leakage power. This makes SRAM less suitable for resource-constrained edge applications where energy efficiency and standby power consumption are critical concerns. Emerging non-volatile memories offer compelling advantages for edge devices, including low standby power consumption and high integration density (thus lower cost per bit) at mature nodes. However, IR drop on bit lines compromises RRAM-ACIM inference accuracy. Furthermore, conventional CIM structures primarily employ voltage or pulse-width modulation for multi-bit WL input in a single cycle. Both methods are susceptible to on-chip noise, degrading inference accuracy as depicted in Fig. 2. Section 3.2 and 3.3 present detailed expositions of the proposed methods N:1 Time Modulation Dynamic

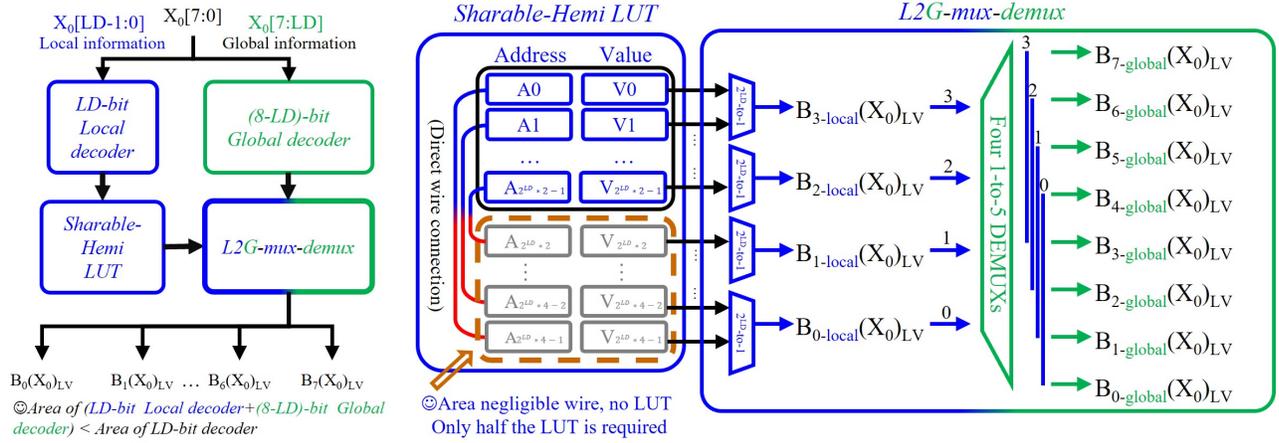

Fig. 5. Hardware architecture with Alignment-Symmetry and PowerGap KAN hardware aware quantization.

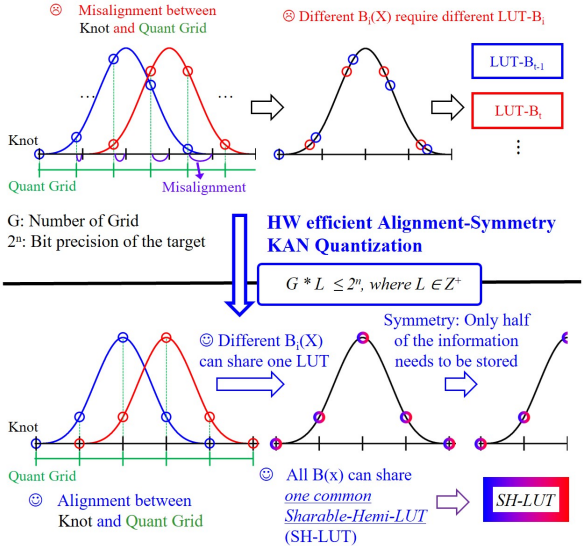

Fig. 3. HW efficient Alignment-Symmetry KAN Quantization for LUTs optimization.

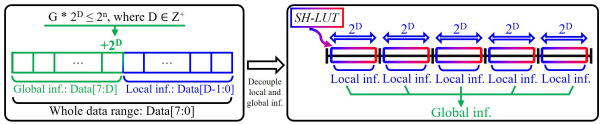

Fig. 4. HW efficient PowerGap KAN Quantization for MUXs and Decoders optimization.

Voltage Input Generator and KAN sparsity-aware weight mapping, which address these challenges comprehensively with RRAM-ACIM.

## 3 PROPOSED TOP-DOWN HW-SW CO-OPTIMIZATION

### 3.1 *Alignment-Symmetry and PowerGap KAN hardware aware quantization for B(X)*

The proposed Alignment-Symmetry and PowerGap KAN hardware aware quantization (ASP-KAN-HAQ) is aimed at reducing the hardware cost and power for evaluating B(X) in Equation (3). ASP-KAN-HAQ comprises two distinct phases:

- Phase one: Alignment-Symmetry for suppressing the needs of programmable LUT by enabling zero offset between knot grid and quantization grid.
- Phase two: PowerGap for mitigating the needs of decoder and MUX by maintaining knot grid spacing at powers of two.

The following analysis considers a scenario where K=3 and G=5, resulting in 8 $B_{0\sim7}(x)$ for each input. We assume 8-bit input precision, with inputs ranging from 0 to 255. However, note that ASP-KAN-HAQ can be generalized to accommodate arbitrary K, G and precision values, as well as layers with negative inputs.

#### A. Phase One: Alignment-Symmetric

The first phase, termed Alignment-Symmetric, is based on the observation illustrated in Fig. 3. The misalignment between knot grid and quantization grid impedes the sharing of a single LUT among different B(X) functions, even when data from various knot grid intervals are shifted to the same interval. To address this issue, ASP-KAN-HAQ guarantees alignment between the knot grid and quantization grid for each B(X) by constraining the quantization grid as an integer multiple of the knot grid, which can be expressed mathematically as:

$$G * L \leq 2^n, \text{where } L \in Z+ \quad (4)$$

Let G denote the number of knots and n the system's maximum bit-width in Equation (4). The value of L that satisfies Equation (4) constrains the data range to the interval [0, G*L-1]. Any L satisfying this integer multiple constraint eliminates the offset between the knot grid and quantization grid, enabling all B(X) functions to share a single LUT. Furthermore, this constraint ensures symmetry in the quantized B(X), which further enables a 50% reduction in the shared LUT size. We refer to this optimized structure as a Sharable-Hemi LUT (SH-LUT).

After Alignment-Symmetric phase, a straightforward approach to route values from the SH-LUT to the corresponding B0~7(x) with minimal hardware employs eight 2L-to-1 transmission gate MUXs (TG-MUXs) and an 8-bit optimized decoder. However, this method still incurs significant area overhead and power consumption.

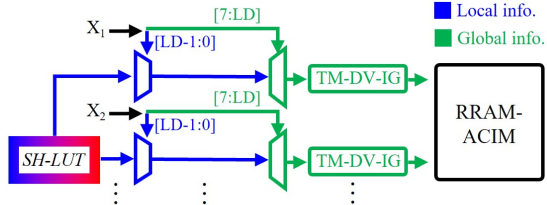

Fig. 6. The hardware architecture with efficient LUT retrieval process.

### B. Phase Two: PowerGap

The second phase, termed PowerGap is proposed to mitigate the TG-MUXs and decoders requirements after Alignment-Symmetric phase. By setting the interval between knot grids to powers of 2, it decouples local and global information, dramatically reducing decoder and TG-MUX areas as shown in Fig. 4, which can be expressed mathematically as:

$$G * 2^D \leq 2^n, \text{where } D \in Z+ \quad (5)$$

We define the information within each knot grid in KAN as local information, while different grid intervals correspond to different B(X), which we define as global information.

The hardware requirements are significantly reduced:
1. TG-MUXs: from original eight 2L-to-1 TG-MUXs to optimized four L-to-1 TG-MUXs and four 1-to-5 TG-DEMUXs.
2. Decoders: from original one 8-bit decoder to optimized one (8-D)-bit decoder and one D-bit decoder.

Given that decoder area grows exponentially with bit width, the area of one 8-bit decoder far exceeds that of one (8-D)-bit decoder plus one D-bit decoder. Therefore, the values that satisfy both Equation (4) and Equation (5) can maximally reduce the area of LUTs, decoders, and TG-MUXs, which can be expressed mathematically as:

$$G * 2^{LD} \leq 2^n, \text{where } LD \in Z+ \quad (6)$$

We designate this optimal value as LD and this value constrains the data range to the interval [0, G*$2^{LD}$-1]. Fig. 5 illustrates the hardware and dataflow for lookup B(X) after optimization using ASP-KAN-HAQ. Fig. 6 shows the efficient lookup process wherein multiple Xi values share a single SH-LUT, facilitating the transfer of corresponding LUT values ($B_{0\sim7\text{-global}}$ (Xi)$_{LV}$) from local to global scope, which are subsequently propagated to the input generator.

### 3.2 N:1 Time Modulation Dynamic Voltage Input Generator for ACIM for $\sum c_i' B_i(X)$

In conventional CIM structures, most multi-bit WL input methods utilize either pure voltage [18][19] or pure pulse-width modulation (PWM) [20][21] to achieve different bit-level WL inputs, which face noise and variation issues or slow down MAC operations. We propose an N:1 Time Modulation Dynamic Voltage Input Generator (TM-DV-IG) that maps distinct lookup table values, representing the corresponding B(X) to multi-level values. This approach distributes the input information across the time and voltage domains, thereby enhancing circuit resistance and maintaining high-speed operation, as depicted in Fig. 7 (b). In the voltage domain, a single RRAM cell's current I on BL is relative to the voltage V applied to WL, denoted as $I[x] \propto f(V[x]), x \in [0, 2^N - 1]$, where $f$ is the $I_d$-$V_g$ curve function of MOSFET. In the time domain, the time t for applying on

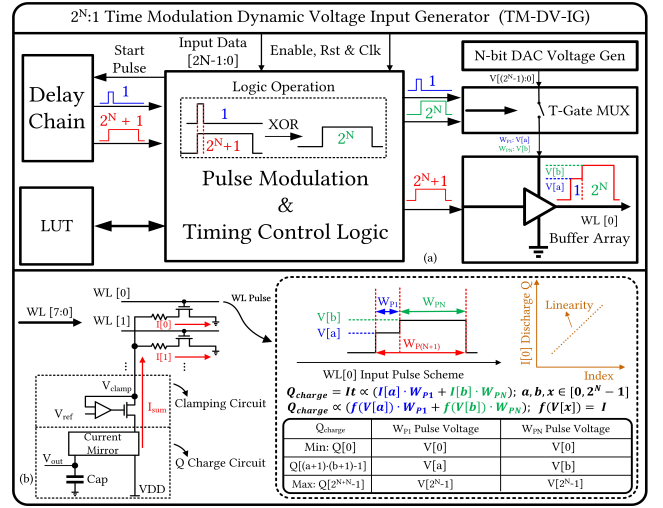

Fig. 7 (a) N:1 Time Modulation Dynamic Voltage input generator for ACIM and (b) BL linear Q value generation theory.

WL is represented by different pulse width W. Ultimately, varying I and t combinations charge different Q into the capacitor. By configuring $I[0]: I[1]: I[2] ... : I[2^N - 1] = 0: 1: 2: ... : 2^N - 1$ by setting the V[$2^N - 1: 0$] properly, the $W_{P1} \cdot I[1]$ can serve as the interval between Q values. This method enables a linear distribution of Q, providing enhanced resistance to noise and variations compared to pure voltage input methods and offering higher operation speed than pure PWM input methods. Consequently, this technique supports the processing of multi-bit inputs within a single clock cycle. The TM-DV-IG comprises five primary components: Delay Chain, Pulse modulation and timing Control Logic (PM-TCM), N-bit DAC, TG-MUX, and buffer array, as shown in Fig. 7.

The PM-TCM generates control signals for switching the buffer array supply voltage and produces input pulses to the buffer array. It collaborates with the delay chain to develop the $1: 2^N: 2^N+1$ ratio pulses ($W_{P1}$, $W_{PN}$, and $W_{P(N+1)}$). The N-bit DAC produces $2^N$ distinct fixed voltage levels and operates with the TG-MUX, which receives $1: 2^N$ ratio pulses generated from PM-TCM. The $2^N+1$ ratio pulse from PM-TCM is supplied to the buffer array, where the buffer output connects with a WL. In the read mode, the TM-DV-IG will transfer the 2N-bit input vector (B(X) values) into a timing modulation dynamic voltage pulse. This pulse is then applied to the WL to generate linearity charge Q, as shown in Fig. 7(a). The corresponding voltage values Vout of charge Q are sensed by the sense amplifier (SA). The CIM architecture allows reusing most circuit blocks for multiple WLs, enabling MD-DV to be extended to the entire array with area-efficient capabilities. For further optimization by co-design ability, we can also optimize the N value for different high-performance (TD-P) and high-accuracy (TD-A) requirements.

### 3.3 KAN sparsity-aware weight mapping for $c_i'$

The parasitic resistance in BLs causes IR-drop, leading to errors in RRAM-ACIM MAC operations that utilize current summation. This, in turn, affects inference accuracy. Previous work [14] has attempted to address this issue; however, these approaches either introduced additional circuitry or imposed limitations on the maximum array size. To address this issue, we propose a KAN sparsity-aware weight mapping technique (KAN-SAM) that requires no modifications to existing hardware or algorithms.

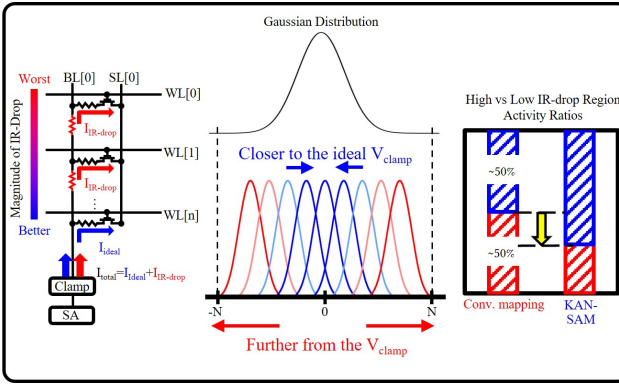
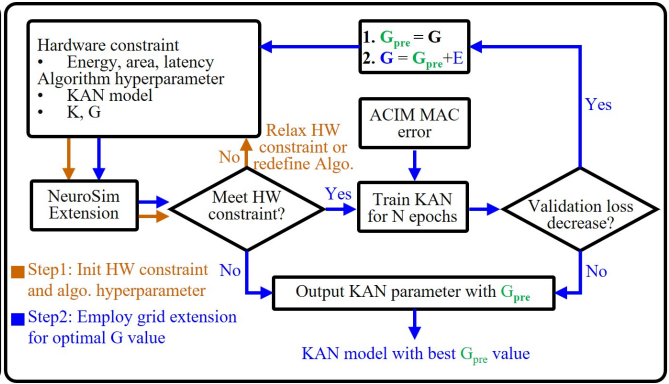

Fig. 8. KAN sparsity-aware weight mapping.

Fig. 9. KAN-NeuroSim hyperparameter optimization framework.

Leveraging the properties of B(X) functions in KAN, not all B(X) are activated for any given input. For instance, when K=3, only four B(X) are triggered simultaneously. By identifying the data range with the highest input probability, we can determine which B(X) has the highest trigger probability. We call this B_H(X). Conversely, the one with the lowest trigger probability is referred to as B_L(X). The ci' coefficients corresponding to B_H(X) can be programmed into RRAM cells closer to the BL clamping circuit, ensuring precision for high-probability inputs. Conversely, ci' coefficients associated with B_L(X) are mapped to cells farther from the clamping circuit, thereby optimizing overall inference accuracy.

Different applications and models exhibit distinct distributions. In Fig. 8, we introduce KAN-SAM using the Gaussian distribution as an example. As illustrated in Fig. 8, for an input range of -N to N, the central Bi(X) function has the highest activation probability, while the Bi(X) functions at the extremes are least likely to be triggered. Consequently, mapping the central ci' values (corresponding to B_H(X)) to RRAM cells nearest to the clamper, and the peripheral ci' values (corresponding to B_L(X)) to more distant cells, maximizes the system's inference accuracy. This principle holds true for input ranges from 0 to N as well.

### 3.4 KAN-NeuroSim hyperparameter optimization framework

Section 3.1 addressed the optimization of B(X) lookup through ASP-KAN-HAQ, generating diverse LD values for varying G parameters. However, a method for determining optimal G values within hardware constraints was not presented. Section 3.2 explored TM-DVS-IG's high-performance (TD-P) and high-accuracy (TD-A) modes. Nevertheless, we lacked an efficient method to analyze the impact of TD-P and TD-A on system performance, enabling users to select the most appropriate mode under different conditions.

To address these limitations, we introduce the KAN-NeuroSim hyperparameter optimization framework (KAN-NeuroSim) in Fig. 9. KAN-NeuroSim comprises two steps. The brown path in Fig. 9 represents step 1, which begins by defining hardware constraints (energy, area, latency) and KAN hyperparameters (network architecture, K, and G). These parameters are then input into the NeuroSim extension [17], which combines ASP-KAN-HAQ and TM-DV-IG to calculate the required energy, area, and latency. If hardware constraints are exceeded, the framework refines either the hardware constraints or the KAN hyperparameters until the requirements are satisfied. Upon meeting hardware constraints, step 2 commences, implementing the grid extension method from the original KAN paper to improve accuracy. During training, the grid is expanded every N epoch. During the training process, G is incrementally incremented by E (user-defined) if the validation loss continues to decrease and the resulting hardware requirements, as assessed by NeuroSim, remain within specified constraints. Otherwise, grid extension terminates, reverting to the previous $G_{pre}$ value. We incorporate RRAM non-ideal effects, specifically partial sum errors, which have been evaluated using statistics measured from TSMC 22nm RRAM-ACIM prototype chips. This integration ensures that the derived KAN hyperparameters achieve optimal hardware efficiency and accuracy when implemented on RRAM-ACIM platforms.

## 4 EVALUATION RESULTS

### A. ASP-KAN-HAQ

We evaluated ASP-KAN-HAQ using the original KAN paper's application in Knot theory, employing a two-layer KAN with dimensions 17x1x14. As discussed in Section 3.1, the value of G is a key parameter for ASP-KAN-HAQ. To quantitatively assess the scalability of our method to more complex KAN architectures with arbitrary G, we incrementally increased G, following the grid extension, a methodology of the original KAN paper authors. We compared ASP-KAN-HAQ with conventional quantization methods [15] in terms of energy and area in 22nm. In this paper, we employ PACT quantization [16] as our baseline for comparison. To isolate variables, our analysis focused on the hardware path spanning from the input X, through the retrieval of the corresponding B(X) value from the LUT, to its subsequent transmission to the input generator. Fig. 10 demonstrates the effectiveness of our method. As G increased from 8 to 64, our approach demonstrates significant improvements over conventional methods, achieving an average area reduction of

40.14x and an average energy reduction of 5.59x. This is attributed to the limitations of conventional quantization methods, where non-zero offsets between quantization grids and knot grids make LUT sharing among different B(X) values challenging. Conversely, our method ensures all B(X) values share a single LUT and decouples local and global information to further reduce TG-MUX and decoder areas ensuring KAN scalability at the edge with ASP-KAN-HAQ.

### B. N:1 Time Modulation Dynamic Voltage input generator

To quantify the benefits of the TM-DV-IG for KAN accelerator implementation, we compared its performance with traditional pure voltage input and PWM methods for multi-bit WL input in Fig. 11, using a 6-bit benchmark requiring $2^6$ distinct pulses for varying BL sampling results. The unit pulse width is assigned to pure voltages, pure PWM, and our TM-DV-IG methods for latency comparison. The evaluation used 22 nm node and all the circuits modules are verified in SPICE for functionality. Our results show that while the pure voltage input method offers the best latency due to its high-bit DAC, it suffers from the small noise margin and static power, leading to significant drawbacks: 1.96x area overhead, 11.9x power overhead compared to TM-DV-IG. The pure PWM approach exhibits the longest latency, incurring an 8x overhead compared to TM-DV-IG. It also incurs a 1.07x area overhead due to the required long delay chain. The TM-DV-IG method combines voltage and PWM, addressing both the noise margin of multi-bit DAC and the timing constraints of pulse width issues. When evaluating the combined metrics of area, power, and latency, TM-DV-IG demonstrates the highest Figure of Merit (FOM), showing a 3x and 4.1x improvement over pure voltage method and PWM method, respectively. The proposed TM-DV-IG significantly aids the KAN algorithm's realization on the RRAM-ACIM.

### C. KAN sparsity-aware weight mapping

We estimated the IR-drop issue and evaluated the proposed KAN-SAM architecture. Firstly, we refer to TSMC's 22 nm RRAM-ACIM chips measurement results [13] of the single BL IR drop effect in array sizes ranging from 128 to 1024. Secondly, we extracted MAC error rates caused by IR-drop from TSMC's 22 nm RRAM-ACIM chips and trained four KAN models using PyTorch, each sized 17x1x14, with G values of 7, 15, 30, and 60 corresponding to array sizes of 128, 256, 512, and 1024, respectively. The baseline uniformly mapped different ci' to RRAM-ACIM without considering Bi(X) activation probabilities. By incorporating the extracted MAC error rates and varying Bi(X) activation probabilities, we analyzed KAN-SAM's impact on accuracy. Fig. 12 demonstrates that as the array dimensions scale from 128 to 1024, the accuracy improvement by KAN-SAM increases from 3.9x to 4.63x. This trend shows the potential of KAN-SAM in enhancing the scalability of RRAM-ACIM systems.

### D. KAN-NeuroSim hyperparameter optimization framework

We employed KAN-NeuroSim with PyTorch environment to optimize the G value for KAN architecture under various hardware constraints on the original KAN paper's application in Knot theory. In this analysis, baseline utilizes a traditional MLP architecture [22] without any techniques proposed in this work. We explored two architectures using minimal (KAN1) and moderate (KAN2) hardware constraints. Fig. 13 demonstrates that due to the high parameter count in traditional MLP, hardware accelerators based on this architecture consume substantially more resources. Compared to the traditional DNN hardware, our KAN-based accelerators with ASP-KAN-HAQ, TM-DV-IG, and KAN-SAM demonstrate a 51.04x to 77.97x reduction in energy consumption, a 9.28x to 41.78x reduction in area, and a 23.59x to 29.56x reduction in latency, while achieving higher accuracy.

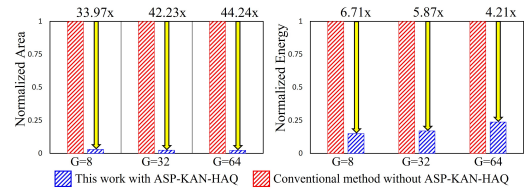

Fig. 10. Comparison of Normalized Area and Energy Consumption between proposed ASP-KAN-HAQ and conventional method based on PACT.

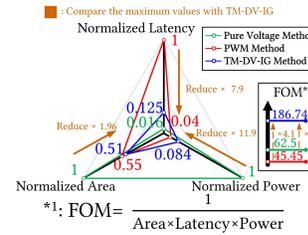
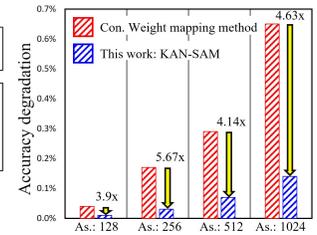

Fig. 11. WL input methods performance comparison with SPICE simulation at 22 nm.

Fig. 12. Comparison of accuracy degradation from KAN software baseline across different RRAM array sizes (As.). The statistics of measured RRAM-ACIM chips [13] are used.

| Metrics | MLP | KAN1 | KAN2 |
|---|---|---|---|
| Area (mm²) | 0.585 | 0.014 | 0.063 |
| Energy (pJ) | 20,049.28 | 257.13 | 392.76 |
| Latency (ns) | 19,632 | 664 | 832 |
| #Param | 190,214 | 279 | 2,232 |
| Accuracy | 78% | 81.03% | 86.74% |

Fig. 13. Comparison of knot theory performance between traditional MLP and optimized KAN accelerators.

## 5 Conclusion

This paper presents, **for the first time,** a novel hardware acceleration approach for KAN via algorithm-hardware co-design. Our proposed algorithm and circuit-level techniques optimize hardware cost, power, and inference accuracy for edge devices. Evaluation results show significant improvements, achieving up to 41.78x reduction in area and 77.97x reduction in energy, with 3.03% accuracy gain compared to MLP approach. Our work addressed the hardware implementation issues of KAN model.

## 6 References

[1] S. Pouyanfar et al., "A survey on deep learning: Algorithms, techniques, and applications," ACM computing surveys, 2019.

[2] W. X. Zhao et al, "A survey of large language models," arXiv:2303.18223, 2023.

[3] Z. Liu et al., "KAN: Kolmogorov-Arnold Networks," arXiv:2404.19756, 2024.

[4] A.N. Kolmogorov, "On the representation of continuous functions of several variables as superpositions of continuous functions of a smaller number of variables," Dokl. Akad. Nauk, 108(2), 1956.

[5] A.N. Kolmogorov, "On the representation of continuous functions of many variables by superposition of continuous functions of one variable and addition," Dokl. Akad. Nauk, Vol. 114. 953–956, 1957.

[6] C. J Vaca-Rubio et al., "Kolmogorov-arnold networks (kans) for time series analysis," arXiv:2405.08790, 2024.

[7] W. J Gordon et al., "B-spline curves and surfaces," In Computer aided geometric design, pages 95–126. Elsevier, 1974.

[8] S. Yu et al., "Compute-in-Memory chips for deep learning: Recent trends and prospects," IEEE Circuits and Systems Magazine, vol. 21, pp. 31-56, 2021.

[9] Y.-D. Chih et al., "An 89 TOPS/W and 16.3 TOPS/mm2 alldigital SRAM-based full-precision compute-in memory macro in 22 nm for machine-learning edge applications," IEEE International Solid-State Circuits Conference (ISSCC), 2021.

[10] X. Si et al., "A Local Computing Cell and 6T SRAM-Based Computing-in-Memory Macro With 8-b MAC Operation for Edge AI Chips," IEEE Journal of Solid-State Circuits (JSSC), vol. 56, no. 9, pp. 2817- 2831, 2021.

[11] J.-W. Su et al., "A 8-b-Precision 6T SRAM Computing-in-Memory Macro Using Segmented-Bitline Charge-Sharing Scheme for AI Edge Chips," IEEE Journal of Solid-State Circuits (JSSC), vol. 57, no. 2, pp. 609–624, 2022.

[12] C.-X. Xue et al., "A 1 Mb multibit ReRAM computing-in-memory macro with 14.6 ns parallel MAC computing time for CNN based AI edge processors," IEEE International Solid-State Circuits Conference (ISSCC), 2019.

[13] W.-H. Huang et al., "A nonvolatile Al-edge processor with 4MB SLC-MLC hybrid-mode ReRAM compute-in-memory macro and 51.4-251TOPS/W," IEEE International Solid-State Circuits Conference (ISSCC), 2023.

[14] B. Liu et al., "Reduction and IR-drop compensations techniques for reliable neuromorphic computing systems," IEEE/ACM International Conference on Computer-Aided Design (ICCAD), 2014.

[15] A. Gholami et al., "A survey of quantization methods for efficient neural network inference," arXiv 2021.

[16] J. Choi et al., "Pact: Parameterized clipping activation for quantized neural networks," arXiv:1805.06085, 2018.

[17] X. Peng et al., "DNN+NeuroSim: An end-to-end benchmarking framework for compute-in-memory accelerators with versatile device technologies," IEEE International Electron Devices Meeting (IEDM), 2019.

[18] Z. Jiang et al.,"C3SRAM: An In-Memory-Computing SRAM Macro Based on Robust Capacitive Coupling Computing Mechanism," IEEE Journal of Solid-State Circuits (JSSC), vol. 55, no. 7, pp. 1888-1897, 2020.

[19] A. Biswas et al., "Conv-RAM: An energy-efficient SRAM with embedded convolution computation for low-power CNN-based machine learning applications," IEEE International Solid-State Circuits Conference - (ISSCC), 2018.

[20] Q. Dong et al., "15.3 A 351TOPS/W and 372.4GOPS Compute-in-Memory SRAM Macro in 7nm FinFET CMOS for Machine-Learning Applications," IEEE International Solid-State Circuits Conference (ISSCC), 2020.

[21] S. K. Gonugondla et al.,"A 42pJ/decision 3.12TOPS/W robust in-memory machine learning classifier with on-chip training," IEEE International Solid-State Circuits Conference (ISSCC), 2018.

[22] A. Davies et al. Advancing mathematics by guiding human intuition with ai. Nature, 600(7887):70–74, 2021.